**The Brightness of Starlink and OneWeb Satellites**

**During Ingress and Egress from Terrestrial Eclipses**

Anthony Mallama

anthony.mallama@gmail.com

2021 December 13

Abstract

A model that combines celestial geometry and atmospheric physics is used to calculate the dimming of artificial satellites as they enter and exit the Earth's shadow. Refraction of sunlight by the terrestrial atmosphere can illuminate a satellite while it is inside the eclipse region determined from geometry alone. Meanwhile, refraction combines with atmospheric absorption to dim the satellites for tens of km outside of that region. Spacecraft brightness is reduced more in blue light than in red because absorption of sunlight is stronger at shorter wavelengths. Observations from the MMT-9 robotic observatory are consistent with the model predictions. Tables of satellite brightness as functions of their location in the eclipse region are provided.

## 1. Introduction

The impact of artificial satellites on astronomical observation has been well documented (Hall et al., 2021, Walker et al., 2020a, Walker et al., 2020b, Tyson et al., 2020, Otarola et al., 2020, Gallozzi et al., 2020, Hainaut and Williams, 2020, McDowell, 2020, Williams et al. 2021a, Williams et al. 2021b). The most numerous of these objects are the Starlink and OneWeb spacecraft.

Several observing programs are underway to record and characterize the brightness of satellites (for example, Tregloan-Reed, 2021). One aim of these studies is to inform astronomers who are scheduling telescope operations so that they can avoid the brightest spacecraft. Another goal is to inform analysts who are developing software that will mitigate the adverse effect of satellite trails on images and spectra.

Much of the research effort so far has centered on measuring the average magnitudes of the satellites (Otarola et al. 2021 and Krantz et al. 2021). Some progress has also been made in deriving equations to represent brightness as a function of the geometry between satellite, Sun and observer (Mallama, 2021).

One aspect of satellite brightness that has been overlooked to date, though, is the eclipse phenomenon. In this paper the term *eclipse* is synonymous with *ingress or egress from the terrestrial shadow*. The analysis centers on ingress and egress where satellite brightness is changing. Spacecraft deep in the shadow are completely invisible and are not discussed.

Sky maps of satellite brightness such as Figure 1 from Mallama (2021) indicate that Starlink VisorSat spacecraft are brightest when they are adjacent to the eclipse region. So, their magnitude changes dramatically over a short distance.

This sensitivity of satellite brightness to position at the eclipse boundary implies that an understanding the ingress/egress phenomenon is important for the mitigation of their adverse effects. A model that calculates the satellite's brightness as a function of its position in the shadow is described in this paper. Tables of intensities and magnitudes in the eclipse are provided.

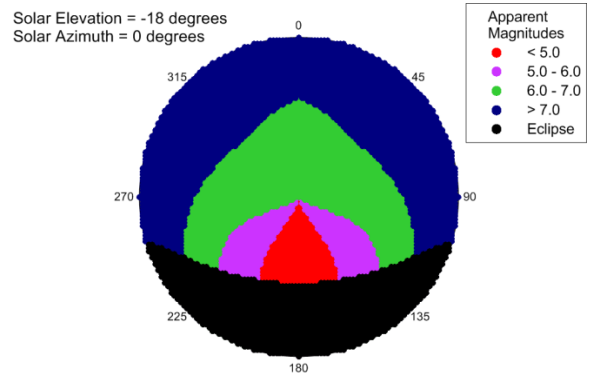

Figure 1. This map of the sky shows the predicted apparent magnitudes of VisorSat satellites according to their elevation above the horizon and their azimuth relative to that of the Sun. Notice that the region with the brightest satellites touches the eclipse area.

Section 2 examines the geometry of the satellite, Sun and Earth during an eclipse. These events occur along a line from the Sun which passes tangentially to the Earth's limb and extends to the spacecraft. Section 3 begins the analysis of atmospheric effects by specifying how air density changes with altitude. Section 4 investigates atmospheric absorption and the transmission of sunlight. Section 5 addresses atmospheric refraction which bends the light rays and also dilutes their intensity. Section 6 describes how geometry, transmission and refraction are combined into an eclipse model. This model is compared with observations in Section 7. Some limitations of this study are addressed in Section 8. Section 9 discusses how to make use of the model results. The conclusions are given in Section 10.

## 2. Satellite-Earth-Sun geometry

A satellite is eclipsed when the Earth's limb, the Sun and the satellite lie on a straight line as illustrated in Figure 2. The Sun's parallel rays are seen entering the diagram from the left. The orange line from the Sun indicates the geometrical eclipse condition and it contacts the Earth's limb tangentially, so it is called the *solar tangent line*.

The diagram also shows some atmospheric effects that are discussed later in the paper. The *minimum altitude* of a ray from the Sun is that height above the limb where it most closely approaches the Earth's surface. The discussions of atmospheric effects in the following sections take heights to be above sea level. So the limb, the tangent line, and the minimum altitude all refer to sea level.

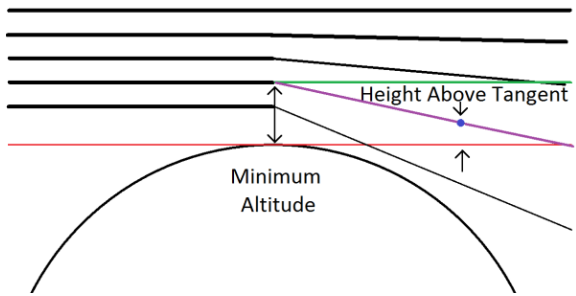

Figure 2. This schematic diagram illustrates eclipse geometry and the refraction of sunlight in the terrestrial atmosphere.

The satellite is represented by the small blue circle on the purple line. That line indicates the ray which passes through a minimum altitude and is refracted toward the Earth center. The green line is the extension of the purple ray but without refraction. The *angle of refraction* is measured between the purple line and the green line. The vertical distance of the satellite from the orange line is the *height above the tangent line*.

Finally, two aspects of eclipse geometry not shown in the Figure are the distance and diameter of the Sun. The effect of those quantities on eclipses is delayed until Section 6 which explains the complete model.

## 3. Atmospheric density

The terrestrial atmosphere above the eclipsing limb absorbs and refracts sunlight on its way to the satellite. Both of these effects depend on atmospheric density, so this section addresses that density as a function of altitude. The U.S. Standard Atmosphere model (Engineering Toolbox, 2003) tabulates the relationship between density and altitude above sea level which is plotted in Figure 3.

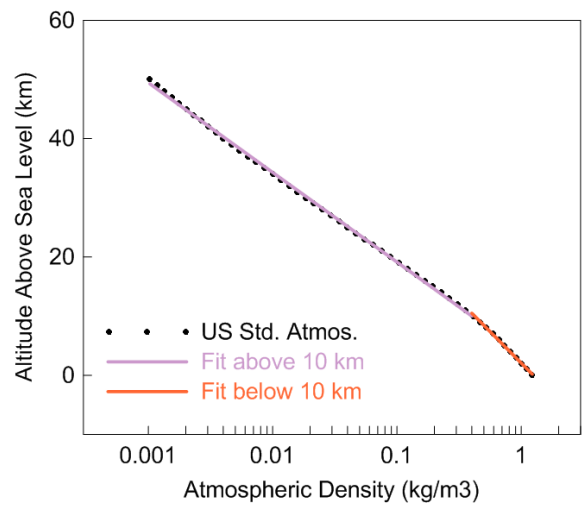

Figure 3. Atmospheric density (X-axis) is plotted as a function of altitude above sea level (Y-axis).

The function begins with 1.225 kg/m$^3$ at sea level and decreases by more than 1000 times at 50 km altitude where it is 0.001027 kg/m$^3$. The Figure indicates that the logarithm of density can be fit with two linear functions which join at 10 km altitude. That height is approximately at the boundary between the troposphere and the stratosphere. The best fit lines are indicated by

Equations 1 and 2 for altitudes above 10 km and below 10 km, respectively,

$$A = -6.583 * \ln(\rho) + 3.973$$

Equation 1

$$A = -9.209 * \ln(\rho) + 2.057$$

Equation 2

where $\rho$ is density in kg/m$^3$ and $A$ is altitude in km. The absolute values of the coefficients of $\ln(\rho)$, 6.583 and 9.209, are the atmospheric *scale heights* in these two regions. Atmospheric density decreases by a factor of $e$ for each scale height of altitude change.

## 4. Absorption and transmission

Rayleigh scattering in the atmosphere reduces the intensity of sunlight on its way to the satellite. The amount light transmitted depends on how much air the ray passes through. The air encountered by a vertical ray that reaches sea level is taken to be *one air mass*.

The corresponding quantity for sunlight passing horizontally through the atmosphere in an eclipse can be computed from the function of density versus altitude above sea level plotted in Figure 3. The cumulative air masses for these horizontal rays are plotted in Figure 4 as a function of the minimum altitude attained by the ray. Those cumulative quantities range from about 0.5 air mass at 50 km minimum altitude to 60 air masses at sea level.

The amount of sunlight transmitted through the atmosphere is given by Equation 3,

$$i(t) / i(0) = E^m$$

Equation 3

where *i(t)* is the intensity of light after passage through the atmosphere, *i(0)* is the intensity of light before passage (so, *i(t)/i(0)* is the relative intensity), *E* is the value of transmission for a unit air mass and *m* is the cumulative air mass.

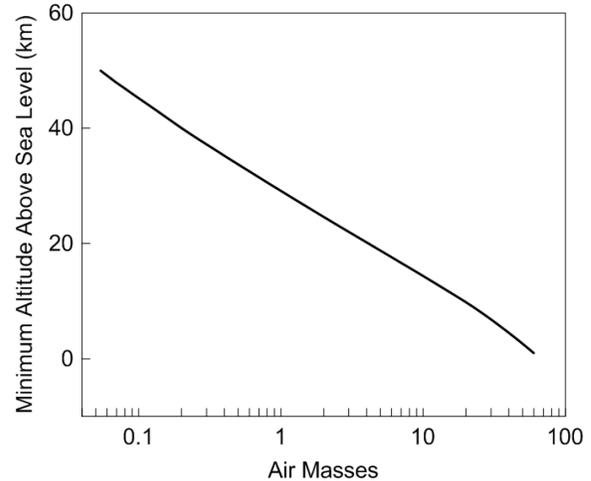

Figure 4. The cumulative air mass encountered by a ray passing horizontally through the atmosphere (X-axis) is plotted as a function of its minimum altitude above sea level (Y-axis).

The value of *E* in Equation 3 depends strongly on wavelength. These values have been measured countless times by astronomers performing stellar photometry and the values adopted in this paper are listed in Table 1. Those entries correspond to color filters in the Johnson and Sloan photometric systems.

Table 1. Values of E (transmission)

| --- Johnson --- | | | ---- Sloan ---- | | |
|---|---|---|---|---|---|
| | microns | E | | microns | E |
| U | 0.36 | 0.50 | u' | 0.36 | 0.50 |
| B | 0.44 | 0.73 | g' | 0.47 | 0.75 |
| V | 0.55 | 0.83 | r' | 0.62 | 0.86 |
| R | 0.70 | 0.90 | i' | 0.75 | 0.91 |
| I | 0.90 | 0.95 | z' | 0.89 | 0.95 |

Color-dependent transmission as a function of minimum altitude was determined by combining the cumulative air masses plotted in Figure 4 with the transmission coefficients listed in Table 1 according to Equation 3. The resulting transmission functions for blue through near-IR wavelengths are shown in Figure 5. Transmissions are all near unity above 50 km and they are all nearly zero at sea level. However, redder wavelengths are much more highly transmitted at intermediate altitudes.

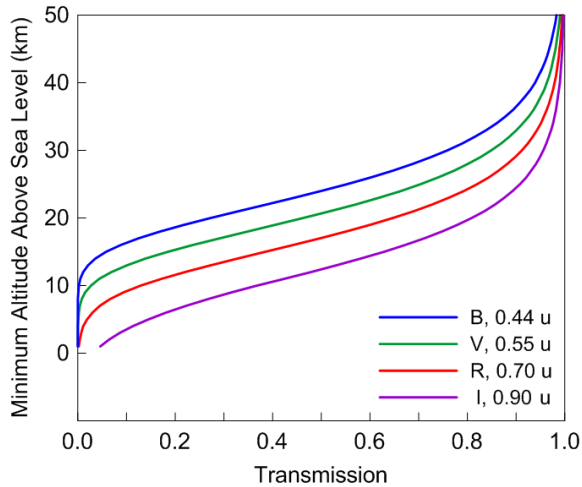

Figure 5. The transmission of sunlight (X-axis) is plotted as a function of the ray's minimum altitude above sea level (Y-axis).

## 5. Refraction

Sunlight passing through the atmosphere is bent toward the center of the Earth by refraction as shown in Figure 2. Refraction grows progressively stronger at lower altitudes where atmospheric density is greater, so differential bending dilutes the intensity of the light. The relationships between atmospheric density, refraction and intensity were originally derived in order to measure the densities of planetary atmospheres from photometry of stellar occultations. These functions can be inverted to model satellite eclipses because the density of the Earth's atmosphere is already known.

The method for computing the effect of refraction on the brightness of an eclipsed satellite is more involved than is that for transmission. There are two positional metrics, namely, the minimum altitude above sea level and the height above the solar tangent line. The former is related to the amount of bending and the latter determines intensity. These two effects are discussed immediately below. Then the difference between the two positional metrics is resolved.

The first of two refraction equations pertains to the minimum altitude for the refracted ray whose intensity is reduced by one-half when it impacts the satellite. That ray is the zero point for computing the geometry and the intensities of all other rays. The minimum altitude for this half-intensity ray can be determined from Equation 4 which is taken from Hunten and Veverka (1976),

$$\theta = H \,/\, D = v \,( 2 \,\pi \,a \,/\, H )^{1/2}$$

Equation 4

where $\theta$ is the refracting angle, $H$ is the atmospheric scale height, $D$ is distance to the satellite, $v$ is index of refraction of the atmosphere minus one, and $a$ is the Earth's radius.

For the distance of a Starlink eclipse from the Earth's limb, 2704 km, and for the scale height of 6.583 km, $v$ is 3.12E-5. Meanwhile, at sea level $v$ = 2.93E-4. Refraction is proportional to density. So, the ratio between these values of $v$ indicates that the ray's minimum altitude occurs where atmospheric density is 0.106 of that at sea level. Solving the density versus altitude function of Equation 1 shows that the minimum altitude of the half-intensity ray above sea level for Starlink satellites is 17.6 km. For OneWeb at an eclipse distance of 4090 km the minimum altitude is found to be 20.2 km.

The height of this half-intensity refracted ray above the solar tangent line at the distance of the satellite is equal to the ray's minimum altitude above sea level minus the product of the refracting angle, $H \,/\, D$, times the distance to the satellite. However, $D$ in the refracting angle cancels *D* for the satellite distance, which just leaves *H*. So, the height above the tangent line is the ray's minimum altitude above sea level minus the scale height. The resulting heights are 11.0 km for Starlink and 13.6 km for One Web. These heights are indicated by the circle symbols in Figure 6. (An alternate method for deducing these half-intensity heights is described in Appendix A.)

The relationship between atmospheric density, refraction angle and minimum altitude above sea level can be used to determine the height above the tangent line for any ray. The resulting function of height above the tangent line versus minimum altitude above sea level is indicated in Figure 6. This correspondence is needed to relate dimming from refraction (which depends on height above the tangent line as explained immediately below) to dimming from atmospheric absorption (which depends on minimum altitude above sea level as explained in the previous section). These two effects are then combined in the eclipse model.

Equation 5 from Wasserman and Veverka (1973) is the second refraction equation. That formula expresses the relationship between the intensity of refracted light and the height above the tangent line at which the ray impacts the satellite,

$$((i(0) / i(r)) - 2) + \ln((i(0) / i(r)) - 1) = \delta(0) - \delta$$

Equation 5

where *i(0)* is the full intensity of light, *i(r)* is the intensity after refraction at the satellite distance (so *i(r)/i(0)* is relative intensity), *δ(0)* is the height of the half-intensity ray measured in scale heights above the tangent line, and *δ* is the distance above the tangent line to the point in question which is also measured in scale heights. Note that solving for the half intensity ray, i(r)/i(0) = 0.5, gives δ(0) − δ = 0. Figure 7 plots this relationship.

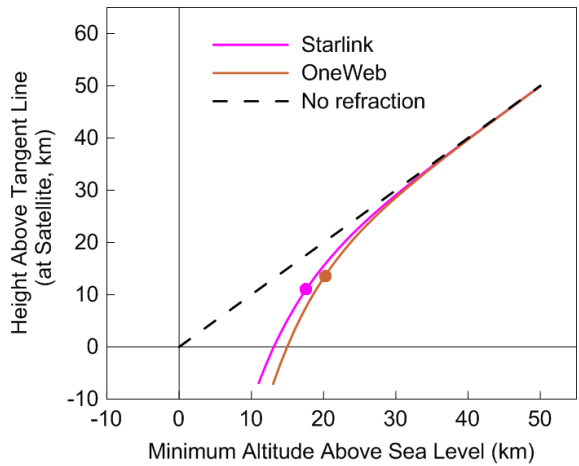

Figure 6. A ray from the Sun passing at the minimum altitude above sea level shown on the X-axis will strike the satellite at the height above the solar tangent line shown on the Y-axis. The two circle symbols are for the rays reduced to half-intensity.

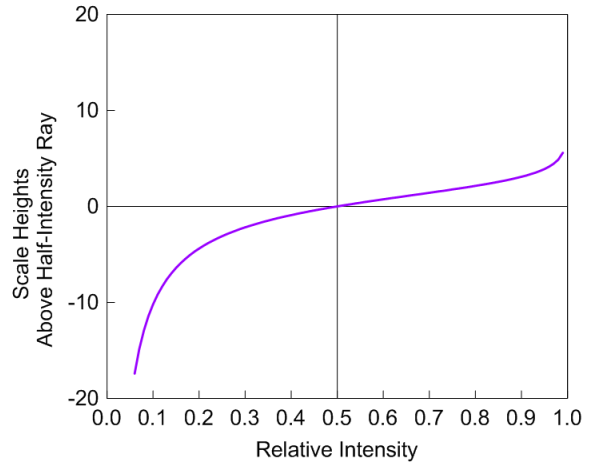

Figure 7. The reduced intensity of light due to refraction (X-axis) is plotted as a function of scale height above the half-intensity ray (Y-axis).

## 6. Eclipse model

The eclipse model uses the product of the transmitted intensity and the refracted intensity of light as indicated by Equations 3 and 5. One complication in taking this product is that these two factors relate to different independent variables. The former is a function of the ray's minimum altitude above sea level, while the latter is a function of height above the half-intensity ray (or above the tangent line). To resolve this difference, the ray's minimum altitude is mapped to its height above the tangent line as indicated in Figure 6. The resulting product of transmission intensity and refraction intensity is the eclipse brightness which is taken as a function of height above the tangent line.

The quantity described above is for a point source of light. However, the Sun has an apparent radius of about one-quarter degree at the Earth's distance, so its size must also be taken into account. The Sun's projected radius at the satellite distances can be computed from its actual radius, its distance from the Earth, and the distances from the limb of 2704 and 4090 km for eclipses of Starlink and OneWeb satellites. Given that input, the Sun's radius projects to 12.6 km and 19.1 km, respectively at the eclipse distances.

Furthermore, the solar disk is not uniformly bright. The term *limb darkening* refers to the decrease in intensity with distance from the center of the solar disk. Equation 6 , from Moon et al. (2017), represents that brightness distribution,

$$i(\phi)1 / i(0) = 1 - \mu (1 - \cos (\phi) )$$

Equation 6

where $\phi$ is the angle measured between a vector from the center of the Sun to the observer and a vector from the center to a location on the photosphere, and where $\mu = 0.69$ is the solar limb darkening coefficient.

A grid of limb-darkened solar brightness was generated and then it was convolved with the product of refracted and transmitted light intensity for a point source. Figure 8 shows the resulting eclipse magnitude function (or light curve) for Starlink in four wavelength bands.

Dimming is greater for shorter wavelengths because more light is absorbed. The graph also indicates dimming of the satellite for the case of 'refraction only', that is, no atmospheric absorption.

Figure 8 additionally plots the 'no atmosphere' case where eclipse brightness is computed only from the geometry of the Sun and Earth. Comparison with the graphs for the color filters shows that the principal effect of the atmosphere is to make the satellite fainter at all heights above the solar tangent line.

When Starlink is near the tangent line it is greatly dimmed in all bands. However, a small amount of light penetrates below that line due to refraction. Meanwhile, at 50 km above the tangent line there is almost no dimming.

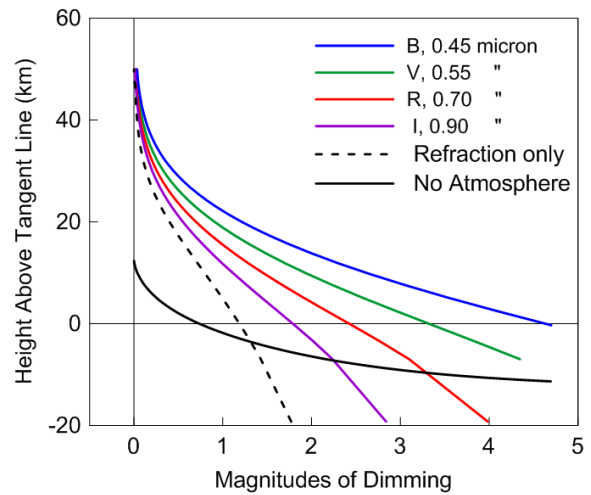

Figure 8. The dimming of a Starlink satellite in eclipse (X-axis) is shown for four wavelength bands as a function of the spacecraft height above the solar tangent line (Y-axis). In the case of 'refraction only', atmospheric transmission is set to unity. 'No atmosphere' indicates dimming that would occur for Sun-Earth-satellite geometry alone, with no refraction or absorption.

The light curve for OneWeb is similar to that of Starlink. However, OneWeb is slightly brighter at small distances above the tangent line because refraction occurs at higher minimum altitudes above sea level where there is greater atmospheric transmission. Meanwhile, OneWeb is slightly fainter than Starlink at large distances above the tangent line because its dimming due to refraction is stronger at a higher altitudes.

Tables of satellite intensity and magnitude for Starlink and OneWeb as functions of their distances above the tangent line are provided in Appendices B and C.

The eclipse model described above pertains to an idealized Earth whose surface is at sea level and where the atmosphere is clear. The eclipse light curve for the real Earth and its changing atmosphere depends on the terrain and on atmospheric conditions at the tangent point. For example, high mountains or clouds can block rays that would otherwise pass close to sea level. Figure 9 shows the effect of 5 km of obstruction. This V-band light curve grows fainter because of blockage as the height above the tangent line decreases and becomes negative.

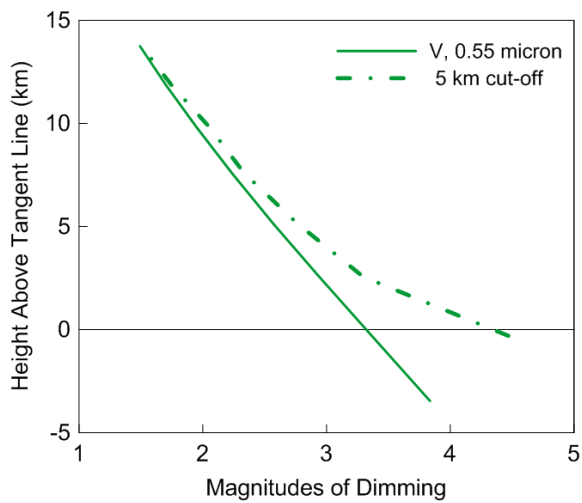

Figure 9. A Starlink eclipse V-band light curve with cut-off 5 km above sea level is compared to the same light curve with no cut-off.

## 7. Comparison with observations

The MMT-9 robotic observatory in Russia records satellite magnitudes in visible light (Karpov et al. 2015 and Beskin et al. 2017). These data have been used to characterize the brightness of Starlink satellites. Mallama (2021) describes the general methods of data processing.

For this study, SGP4 software was used to compute the satellite positions relative to the Earth. The position of the Sun was taken from the JPL Horizons ephemeris system. Then the satellite height above the solar tangent line was determined. The radius of MMT-9 from the geocenter is 6368 km and an additional 2 km was added to account for high terrain in the region and the possibility of clouds. This resulted in 6370 km as the radius used for computations. Apparent magnitudes were adjusted to a standard distance of 1000 km. Those standardized magnitudes for each dataset were normalized to the satellite brightness of that pass at 50 km above the tangent line where it should be near full luminosity. Thus the magnitudes at lower heights reflect dimming due to the eclipse.

The MMT-9 database contains *track* files which record all the data from one satellite pass. Ten track files with eclipses were located and that data was processed as described above. Starlink satellites exhibit many sudden brightness changes even outside of eclipses. Two of the 10 passes showed such irregularities and were omitted from further analysis. Appendix D lists all the tracks that were retained and shows the irregular light curves for the two that were rejected.

Figure 10 displays a composite light curve of the MMT-9 data. The general trend of the observations is consistent with the V-band model light curve. There are differences of up to a magnitude but those data are both brighter and fainter than the model. By contrast the observed magnitudes are systematically fainter than the 'no atmosphere' model shown in the Figure.

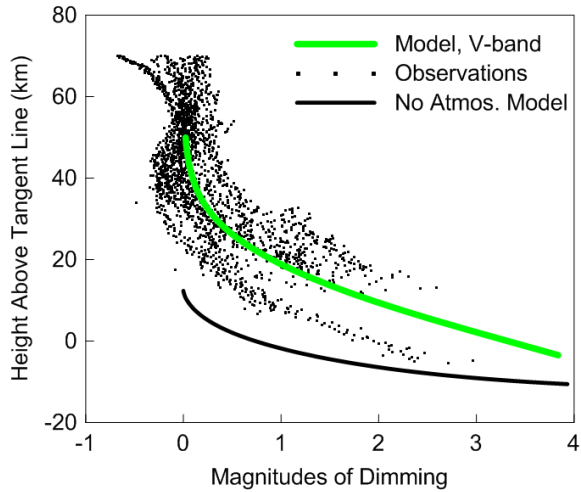

Figure 10. Visible light magnitudes from MMT-9 are compared to the eclipse model for the V-band. The observations fit that model much more closely than the 'no atmosphere' model.

## 8. Limitations of this study

Some simplifications were employed during the development of the model. For example, the distance of an eclipsed satellite from the point of contact between the solar tangent line and the Earth limb is taken to be 2704 km for Starlink and 4090 km for OneWeb. Those values are measured along the solar tangent line, so satellites above or below that line will really be at slightly different distances. However, the model uses the constant distance values.

Another approximation pertains to solar limb-darkening which actually depends on the wavelength of light. Only one representative limb-darkening coefficient was used to model the brightness distribution of the Sun in this study.

Finally, the coefficients of atmospheric transmission listed in Table 1 are those which are familiar to the author from practical experience with astronomical photometry.

There may be other values that are more representative.

## 9. Using the model results

The results of this study may be useful to astronomers who are planning or analyzing observations. Brightness here is expressed as a function of height above the solar tangent line. The computation of that height is described next so that the Tables in Appendices B and C may be used.

The tangent line lays along a cone that fits the Earth's limb and whose apex is at the center of the solar disk. The cone expands at a rate of just 4E-5 km for each km of eclipse distance. So it may be considered to be a cylinder with the Earth's radius at eclipse distances out to a few thousand km with an error that is less than one km. The axis of the cylinder is on a line that connects the Sun and Earth. The height above the solar tangent line, then, is the distance of a satellite from the axial line minus the adopted Earth radius.

When the height of a satellite above the tangent line is known, the dimming due to eclipse can be read from the Tables in Appendices B and C.

## 10. Conclusions

The geometry of the Sun, Earth and satellite is combined with the effects of refraction and absorption in the terrestrial atmosphere to produce a model that predicts dimming of satellites entering or exiting in the Earth's shadow. Refraction and absorption cause dimming up to tens of km above the geometric upper boundary of an eclipse. Likewise, refraction bends some light below the height were geometry alone would block all sunlight. Observed magnitudes are in general agreement with the model. Tables of dimming are provided and the method for using them is discussed.

## Appendix A. An alternate method for computing the height of the half-intensity ray

Section 5 explained how the height of the half-intensity ray above the solar tangent line can be determined from Equation 4. Alternately, that height can be derived from the fact that refraction of a celestial body on the horizon for an observer at sea level is 35' (35 arc-minutes) according to Bennett (1982). Since a light ray passes into and out of the atmosphere for a satellite eclipse, that value of refraction is doubled.

Section 5 also showed that the atmospheric density at the minimum height of the half-intensity ray is 0.106 times that at sea level. Since refraction is proportional to density, the refraction angle for the half-intensity eclipse ray is 0.106 times 70', or 0.00217 radians for Starlink.

When that refraction angle is extended for 2704 km for Starlink satellites and subtracted from the minimum altitude of the ray above sea level, the height above the tangent line is found to be 11.7 km. For OneWeb at 4090 km that height is 14.3 km. These values compare closely to 11.0 and 13.6 km from Section 5.

## Appendix B. Tables of satellite eclipse intensities

The first column is the satellite height measured above the solar tangent line in km. The second column is the dimmed light intensity due to refraction only. The remaining columns are the intensities from the full eclipse model in 10 wavelength bands. Intensities are normalized to unity outside of eclipse.

### Table B.1. Starlink eclipse intensities

| HT_KM | Refr | U | B | V | R | I | u | g | r | i | z |
|---|---|---|---|---|---|---|---|---|---|---|---|
| -40.64 | 0.142 | 0.000 | 0.000 | 0.000 | 0.005 | 0.027 | 0.000 | 0.000 | 0.001 | 0.007 | 0.027 |
| -34.69 | 0.154 | 0.000 | 0.000 | 0.001 | 0.008 | 0.036 | 0.000 | 0.000 | 0.002 | 0.011 | 0.036 |
| -29.15 | 0.167 | 0.000 | 0.000 | 0.002 | 0.012 | 0.046 | 0.000 | 0.000 | 0.004 | 0.016 | 0.046 |
| -24.00 | 0.181 | 0.000 | 0.000 | 0.003 | 0.018 | 0.059 | 0.000 | 0.000 | 0.007 | 0.023 | 0.059 |
| -19.22 | 0.196 | 0.000 | 0.000 | 0.005 | 0.025 | 0.073 | 0.000 | 0.001 | 0.011 | 0.032 | 0.073 |
| -6.97 | 0.268 | 0.000 | 0.003 | 0.018 | 0.058 | 0.128 | 0.000 | 0.005 | 0.030 | 0.068 | 0.128 |
| -3.46 | 0.299 | 0.000 | 0.007 | 0.029 | 0.078 | 0.156 | 0.000 | 0.009 | 0.045 | 0.090 | 0.156 |
| -0.30 | 0.332 | 0.001 | 0.013 | 0.045 | 0.104 | 0.189 | 0.001 | 0.017 | 0.065 | 0.118 | 0.189 |
| 2.56 | 0.366 | 0.002 | 0.024 | 0.067 | 0.136 | 0.226 | 0.002 | 0.029 | 0.091 | 0.151 | 0.226 |
| 5.16 | 0.402 | 0.004 | 0.039 | 0.094 | 0.173 | 0.267 | 0.004 | 0.046 | 0.122 | 0.189 | 0.267 |
| 7.59 | 0.440 | 0.009 | 0.060 | 0.128 | 0.215 | 0.311 | 0.009 | 0.070 | 0.160 | 0.231 | 0.311 |
| 9.81 | 0.478 | 0.018 | 0.087 | 0.166 | 0.259 | 0.356 | 0.018 | 0.099 | 0.201 | 0.276 | 0.356 |
| 11.85 | 0.517 | 0.031 | 0.120 | 0.208 | 0.305 | 0.401 | 0.031 | 0.134 | 0.245 | 0.322 | 0.401 |
| 13.75 | 0.555 | 0.048 | 0.157 | 0.253 | 0.352 | 0.445 | 0.048 | 0.173 | 0.292 | 0.369 | 0.445 |
| 15.51 | 0.592 | 0.071 | 0.198 | 0.300 | 0.399 | 0.488 | 0.071 | 0.215 | 0.339 | 0.415 | 0.488 |
| 17.17 | 0.629 | 0.099 | 0.242 | 0.347 | 0.445 | 0.530 | 0.099 | 0.260 | 0.386 | 0.461 | 0.530 |
| 18.73 | 0.664 | 0.131 | 0.287 | 0.394 | 0.489 | 0.570 | 0.131 | 0.306 | 0.433 | 0.505 | 0.570 |
| 20.22 | 0.697 | 0.167 | 0.334 | 0.441 | 0.532 | 0.608 | 0.167 | 0.353 | 0.478 | 0.547 | 0.608 |
| 21.63 | 0.729 | 0.206 | 0.381 | 0.486 | 0.573 | 0.644 | 0.206 | 0.400 | 0.522 | 0.587 | 0.644 |
| 22.98 | 0.758 | 0.247 | 0.427 | 0.529 | 0.612 | 0.677 | 0.247 | 0.446 | 0.563 | 0.624 | 0.677 |
| 24.27 | 0.784 | 0.290 | 0.472 | 0.570 | 0.648 | 0.708 | 0.290 | 0.490 | 0.603 | 0.660 | 0.708 |
| 25.52 | 0.809 | 0.334 | 0.515 | 0.609 | 0.682 | 0.737 | 0.334 | 0.533 | 0.640 | 0.692 | 0.737 |
| 26.73 | 0.831 | 0.378 | 0.557 | 0.646 | 0.713 | 0.764 | 0.378 | 0.574 | 0.674 | 0.723 | 0.764 |
| 27.91 | 0.851 | 0.423 | 0.597 | 0.680 | 0.742 | 0.788 | 0.423 | 0.613 | 0.707 | 0.751 | 0.788 |
| 29.07 | 0.869 | 0.467 | 0.635 | 0.712 | 0.769 | 0.811 | 0.467 | 0.650 | 0.737 | 0.777 | 0.811 |
| 30.20 | 0.885 | 0.510 | 0.670 | 0.742 | 0.794 | 0.831 | 0.510 | 0.684 | 0.764 | 0.801 | 0.831 |
| 31.31 | 0.900 | 0.551 | 0.703 | 0.769 | 0.816 | 0.850 | 0.551 | 0.716 | 0.789 | 0.823 | 0.850 |
| 32.41 | 0.913 | 0.591 | 0.734 | 0.794 | 0.837 | 0.867 | 0.591 | 0.746 | 0.812 | 0.843 | 0.867 |
| 33.50 | 0.924 | 0.629 | 0.762 | 0.817 | 0.855 | 0.882 | 0.629 | 0.773 | 0.833 | 0.861 | 0.882 |
| 34.57 | 0.934 | 0.664 | 0.788 | 0.838 | 0.872 | 0.896 | 0.664 | 0.798 | 0.853 | 0.877 | 0.896 |
| 35.63 | 0.943 | 0.698 | 0.811 | 0.857 | 0.887 | 0.908 | 0.698 | 0.821 | 0.870 | 0.891 | 0.908 |
| 36.68 | 0.950 | 0.729 | 0.833 | 0.873 | 0.901 | 0.919 | 0.729 | 0.841 | 0.885 | 0.904 | 0.919 |
| 37.73 | 0.957 | 0.758 | 0.853 | 0.889 | 0.913 | 0.930 | 0.758 | 0.860 | 0.899 | 0.916 | 0.930 |
| 38.76 | 0.963 | 0.784 | 0.870 | 0.902 | 0.924 | 0.939 | 0.784 | 0.877 | 0.912 | 0.927 | 0.939 |
| 39.80 | 0.968 | 0.808 | 0.886 | 0.915 | 0.934 | 0.947 | 0.808 | 0.892 | 0.923 | 0.936 | 0.947 |
| 40.82 | 0.972 | 0.830 | 0.900 | 0.926 | 0.942 | 0.954 | 0.830 | 0.905 | 0.933 | 0.945 | 0.954 |
| 41.85 | 0.976 | 0.849 | 0.912 | 0.935 | 0.950 | 0.960 | 0.849 | 0.917 | 0.941 | 0.952 | 0.960 |
| 42.87 | 0.979 | 0.868 | 0.924 | 0.944 | 0.957 | 0.966 | 0.868 | 0.928 | 0.950 | 0.959 | 0.966 |
| 43.88 | 0.982 | 0.882 | 0.933 | 0.951 | 0.962 | 0.970 | 0.882 | 0.937 | 0.956 | 0.964 | 0.970 |
| 44.90 | 0.984 | 0.896 | 0.941 | 0.958 | 0.968 | 0.975 | 0.896 | 0.945 | 0.962 | 0.969 | 0.975 |
| 45.91 | 0.986 | 0.908 | 0.949 | 0.963 | 0.973 | 0.979 | 0.908 | 0.952 | 0.967 | 0.974 | 0.979 |
| 46.92 | 0.988 | 0.918 | 0.955 | 0.968 | 0.976 | 0.981 | 0.918 | 0.957 | 0.971 | 0.977 | 0.981 |
| 47.93 | 1.000 | 0.928 | 0.961 | 0.973 | 0.980 | 0.985 | 0.928 | 0.964 | 0.976 | 0.981 | 0.985 |
| 48.94 | 1.000 | 0.934 | 0.965 | 0.976 | 0.982 | 0.987 | 0.934 | 0.967 | 0.979 | 0.983 | 0.987 |
| 49.95 | 1.000 | 0.941 | 0.969 | 0.979 | 0.985 | 0.989 | 0.941 | 0.971 | 0.982 | 0.986 | 0.989 |

## Table B.2. OneWeb eclipse intensities

| HT_KM | Refr | U | B | V | R | I | u | g | r | i | z |
|---|---|---|---|---|---|---|---|---|---|---|---|
| -40.93 | 0.131 | 0.000 | 0.000 | 0.002 | 0.013 | 0.043 | 0.000 | 0.000 | 0.005 | 0.017 | 0.043 |
| -34.20 | 0.142 | 0.000 | 0.000 | 0.004 | 0.018 | 0.053 | 0.000 | 0.001 | 0.008 | 0.023 | 0.053 |
| -16.18 | 0.196 | 0.000 | 0.003 | 0.015 | 0.044 | 0.096 | 0.000 | 0.004 | 0.024 | 0.052 | 0.096 |
| -11.38 | 0.220 | 0.000 | 0.006 | 0.024 | 0.061 | 0.119 | 0.000 | 0.008 | 0.036 | 0.070 | 0.119 |
| -7.11 | 0.248 | 0.001 | 0.012 | 0.038 | 0.083 | 0.147 | 0.001 | 0.015 | 0.053 | 0.093 | 0.147 |
| -3.30 | 0.277 | 0.002 | 0.022 | 0.057 | 0.111 | 0.179 | 0.002 | 0.027 | 0.076 | 0.122 | 0.179 |
| 0.12 | 0.308 | 0.006 | 0.037 | 0.082 | 0.143 | 0.215 | 0.006 | 0.044 | 0.104 | 0.155 | 0.215 |
| 3.28 | 0.342 | 0.012 | 0.058 | 0.113 | 0.181 | 0.255 | 0.012 | 0.067 | 0.138 | 0.194 | 0.255 |
| 6.12 | 0.378 | 0.023 | 0.085 | 0.149 | 0.222 | 0.297 | 0.023 | 0.095 | 0.177 | 0.235 | 0.297 |
| 8.70 | 0.416 | 0.039 | 0.117 | 0.188 | 0.265 | 0.340 | 0.039 | 0.129 | 0.218 | 0.278 | 0.340 |
| 11.05 | 0.454 | 0.059 | 0.153 | 0.230 | 0.309 | 0.382 | 0.059 | 0.166 | 0.261 | 0.322 | 0.382 |
| 13.21 | 0.492 | 0.085 | 0.193 | 0.274 | 0.353 | 0.424 | 0.085 | 0.207 | 0.305 | 0.366 | 0.424 |
| 15.21 | 0.531 | 0.115 | 0.234 | 0.318 | 0.396 | 0.465 | 0.115 | 0.249 | 0.349 | 0.409 | 0.465 |
| 17.06 | 0.569 | 0.148 | 0.277 | 0.362 | 0.439 | 0.505 | 0.148 | 0.292 | 0.393 | 0.451 | 0.505 |
| 18.79 | 0.607 | 0.183 | 0.319 | 0.405 | 0.480 | 0.542 | 0.183 | 0.335 | 0.435 | 0.491 | 0.542 |
| 20.41 | 0.643 | 0.220 | 0.362 | 0.447 | 0.519 | 0.578 | 0.220 | 0.377 | 0.476 | 0.530 | 0.578 |
| 21.94 | 0.677 | 0.258 | 0.403 | 0.487 | 0.556 | 0.611 | 0.258 | 0.419 | 0.515 | 0.566 | 0.611 |
| 23.38 | 0.709 | 0.297 | 0.444 | 0.525 | 0.590 | 0.642 | 0.297 | 0.459 | 0.552 | 0.600 | 0.642 |
| 24.76 | 0.739 | 0.335 | 0.482 | 0.561 | 0.623 | 0.672 | 0.335 | 0.497 | 0.587 | 0.632 | 0.672 |
| 26.08 | 0.768 | 0.374 | 0.520 | 0.595 | 0.654 | 0.699 | 0.374 | 0.534 | 0.620 | 0.663 | 0.699 |
| 27.36 | 0.793 | 0.412 | 0.556 | 0.628 | 0.683 | 0.725 | 0.412 | 0.569 | 0.651 | 0.691 | 0.725 |
| 28.59 | 0.817 | 0.449 | 0.590 | 0.658 | 0.710 | 0.749 | 0.449 | 0.603 | 0.680 | 0.718 | 0.749 |
| 29.79 | 0.839 | 0.486 | 0.622 | 0.687 | 0.735 | 0.771 | 0.486 | 0.635 | 0.708 | 0.742 | 0.771 |
| 30.96 | 0.858 | 0.522 | 0.653 | 0.714 | 0.759 | 0.792 | 0.522 | 0.665 | 0.733 | 0.766 | 0.792 |
| 32.11 | 0.875 | 0.556 | 0.683 | 0.740 | 0.781 | 0.811 | 0.556 | 0.694 | 0.757 | 0.787 | 0.811 |
| 33.24 | 0.891 | 0.590 | 0.710 | 0.764 | 0.802 | 0.829 | 0.590 | 0.721 | 0.780 | 0.807 | 0.829 |
| 34.34 | 0.905 | 0.622 | 0.736 | 0.786 | 0.821 | 0.846 | 0.622 | 0.746 | 0.801 | 0.826 | 0.846 |
| 35.44 | 0.917 | 0.653 | 0.761 | 0.806 | 0.838 | 0.861 | 0.653 | 0.770 | 0.820 | 0.843 | 0.861 |
| 36.52 | 0.928 | 0.682 | 0.783 | 0.825 | 0.855 | 0.875 | 0.682 | 0.792 | 0.838 | 0.859 | 0.875 |
| 37.59 | 0.937 | 0.709 | 0.804 | 0.843 | 0.870 | 0.888 | 0.709 | 0.812 | 0.854 | 0.873 | 0.888 |
| 38.64 | 0.946 | 0.735 | 0.824 | 0.859 | 0.883 | 0.900 | 0.735 | 0.831 | 0.870 | 0.887 | 0.900 |
| 39.69 | 0.953 | 0.759 | 0.841 | 0.874 | 0.895 | 0.911 | 0.759 | 0.848 | 0.883 | 0.899 | 0.911 |
| 40.73 | 0.959 | 0.781 | 0.857 | 0.887 | 0.907 | 0.920 | 0.781 | 0.863 | 0.895 | 0.909 | 0.920 |
| 41.77 | 0.964 | 0.802 | 0.872 | 0.899 | 0.917 | 0.929 | 0.802 | 0.878 | 0.907 | 0.919 | 0.929 |
| 42.80 | 0.970 | 0.821 | 0.886 | 0.910 | 0.926 | 0.937 | 0.821 | 0.891 | 0.917 | 0.928 | 0.937 |
| 43.82 | 0.973 | 0.838 | 0.897 | 0.919 | 0.934 | 0.944 | 0.838 | 0.902 | 0.926 | 0.936 | 0.944 |
| 44.85 | 0.977 | 0.854 | 0.908 | 0.928 | 0.941 | 0.950 | 0.854 | 0.912 | 0.934 | 0.943 | 0.950 |
| 45.86 | 0.980 | 0.869 | 0.918 | 0.936 | 0.948 | 0.956 | 0.869 | 0.922 | 0.941 | 0.949 | 0.956 |
| 46.88 | 0.983 | 0.881 | 0.926 | 0.943 | 0.953 | 0.960 | 0.881 | 0.930 | 0.947 | 0.955 | 0.960 |
| 47.90 | 0.985 | 0.892 | 0.933 | 0.948 | 0.958 | 0.964 | 0.892 | 0.937 | 0.953 | 0.959 | 0.964 |
| 48.91 | 0.987 | 0.902 | 0.940 | 0.954 | 0.962 | 0.968 | 0.902 | 0.943 | 0.957 | 0.964 | 0.968 |
| 49.92 | 0.989 | 0.911 | 0.946 | 0.958 | 0.966 | 0.972 | 0.911 | 0.948 | 0.962 | 0.967 | 0.972 |

## Appendix C. Tables of satellite eclipse magnitudes

The first column is the satellite height measured above the solar tangent line in km. The second column is the dimming in magnitudes from refraction only. The remaining columns are magnitudes of dimming from the full eclipse model in 10 wavelength bands. The magnitudes are normalized to zero outside of eclipse.

## Table C.1. Starlink eclipse magnitudes

| HT_KM | Refr | U | B | V | R | I | u | g | r | i | z |
|---|---|---|---|---|---|---|---|---|---|---|---|
| -40.64 | 2.119 | 23.977 | 12.636 | 8.512 | 5.799 | 3.930 | 23.977 | 11.783 | 7.334 | 5.421 | 3.930 |
| -34.69 | 2.030 | 21.176 | 11.253 | 7.641 | 5.262 | 3.622 | 21.176 | 10.507 | 6.609 | 4.930 | 3.622 |
| -29.15 | 1.943 | 18.596 | 9.989 | 6.844 | 4.768 | 3.334 | 18.596 | 9.340 | 5.943 | 4.478 | 3.334 |
| -24.00 | 1.856 | 16.716 | 8.981 | 6.175 | 4.335 | 3.070 | 16.716 | 8.401 | 5.376 | 4.079 | 3.070 |
| -19.22 | 1.770 | 15.580 | 8.261 | 5.665 | 3.984 | 2.839 | 15.580 | 7.720 | 4.932 | 3.752 | 2.839 |
| -6.97 | 1.428 | 11.090 | 6.212 | 4.346 | 3.099 | 2.236 | 11.090 | 5.830 | 3.806 | 2.925 | 2.236 |
| -3.46 | 1.311 | 9.457 | 5.408 | 3.836 | 2.768 | 2.019 | 9.457 | 5.088 | 3.375 | 2.617 | 2.019 |
| -0.30 | 1.198 | 8.132 | 4.699 | 3.365 | 2.454 | 1.809 | 8.132 | 4.428 | 2.973 | 2.325 | 1.809 |
| 2.56 | 1.091 | 6.934 | 4.071 | 2.941 | 2.164 | 1.613 | 6.934 | 3.842 | 2.607 | 2.054 | 1.613 |
| 5.16 | 0.990 | 5.936 | 3.526 | 2.566 | 1.904 | 1.433 | 5.936 | 3.332 | 2.282 | 1.810 | 1.433 |
| 7.59 | 0.892 | 5.085 | 3.051 | 2.235 | 1.671 | 1.269 | 5.085 | 2.886 | 1.993 | 1.591 | 1.269 |
| 9.81 | 0.801 | 4.380 | 2.647 | 1.950 | 1.467 | 1.123 | 4.380 | 2.507 | 1.742 | 1.398 | 1.123 |
| 11.85 | 0.717 | 3.788 | 2.303 | 1.703 | 1.289 | 0.993 | 3.788 | 2.182 | 1.525 | 1.230 | 0.993 |
| 13.75 | 0.640 | 3.293 | 2.010 | 1.492 | 1.134 | 0.879 | 3.293 | 1.905 | 1.338 | 1.083 | 0.879 |
| 15.51 | 0.569 | 2.874 | 1.759 | 1.309 | 0.999 | 0.779 | 2.874 | 1.668 | 1.176 | 0.955 | 0.779 |
| 17.17 | 0.503 | 2.515 | 1.541 | 1.149 | 0.880 | 0.689 | 2.515 | 1.462 | 1.034 | 0.842 | 0.689 |
| 18.73 | 0.445 | 2.209 | 1.354 | 1.011 | 0.776 | 0.610 | 2.209 | 1.284 | 0.910 | 0.743 | 0.610 |
| 20.22 | 0.391 | 1.946 | 1.190 | 0.890 | 0.685 | 0.540 | 1.946 | 1.129 | 0.801 | 0.655 | 0.540 |
| 21.63 | 0.344 | 1.717 | 1.048 | 0.784 | 0.604 | 0.478 | 1.717 | 0.995 | 0.706 | 0.579 | 0.478 |
| 22.98 | 0.301 | 1.518 | 0.924 | 0.691 | 0.533 | 0.423 | 1.518 | 0.877 | 0.623 | 0.511 | 0.423 |
| 24.27 | 0.264 | 1.344 | 0.816 | 0.610 | 0.471 | 0.375 | 1.344 | 0.774 | 0.550 | 0.452 | 0.375 |
| 25.52 | 0.230 | 1.191 | 0.720 | 0.538 | 0.416 | 0.331 | 1.191 | 0.683 | 0.485 | 0.399 | 0.331 |
| 26.73 | 0.201 | 1.055 | 0.635 | 0.474 | 0.367 | 0.293 | 1.055 | 0.602 | 0.428 | 0.352 | 0.293 |
| 27.91 | 0.175 | 0.935 | 0.560 | 0.418 | 0.324 | 0.259 | 0.935 | 0.531 | 0.377 | 0.311 | 0.259 |
| 29.07 | 0.152 | 0.827 | 0.494 | 0.368 | 0.285 | 0.228 | 0.827 | 0.468 | 0.332 | 0.274 | 0.228 |
| 30.20 | 0.132 | 0.732 | 0.435 | 0.324 | 0.251 | 0.201 | 0.732 | 0.412 | 0.292 | 0.241 | 0.201 |
| 31.31 | 0.114 | 0.647 | 0.383 | 0.285 | 0.221 | 0.177 | 0.647 | 0.362 | 0.257 | 0.212 | 0.177 |
| 32.41 | 0.099 | 0.571 | 0.336 | 0.250 | 0.194 | 0.155 | 0.571 | 0.318 | 0.226 | 0.186 | 0.155 |
| 33.50 | 0.086 | 0.504 | 0.295 | 0.219 | 0.170 | 0.136 | 0.504 | 0.280 | 0.198 | 0.163 | 0.136 |
| 34.57 | 0.074 | 0.444 | 0.259 | 0.192 | 0.149 | 0.119 | 0.444 | 0.245 | 0.173 | 0.143 | 0.119 |
| 35.63 | 0.064 | 0.391 | 0.227 | 0.168 | 0.130 | 0.104 | 0.391 | 0.215 | 0.152 | 0.125 | 0.104 |
| 36.68 | 0.055 | 0.343 | 0.199 | 0.147 | 0.114 | 0.091 | 0.343 | 0.188 | 0.132 | 0.109 | 0.091 |
| 37.73 | 0.048 | 0.301 | 0.173 | 0.128 | 0.099 | 0.079 | 0.301 | 0.164 | 0.115 | 0.095 | 0.079 |
| 38.76 | 0.041 | 0.264 | 0.151 | 0.112 | 0.086 | 0.069 | 0.264 | 0.143 | 0.100 | 0.083 | 0.069 |
| 39.80 | 0.036 | 0.231 | 0.132 | 0.097 | 0.075 | 0.060 | 0.231 | 0.124 | 0.087 | 0.072 | 0.060 |
| 40.82 | 0.031 | 0.202 | 0.115 | 0.084 | 0.064 | 0.051 | 0.202 | 0.108 | 0.075 | 0.062 | 0.051 |
| 41.85 | 0.027 | 0.177 | 0.100 | 0.073 | 0.056 | 0.045 | 0.177 | 0.095 | 0.066 | 0.054 | 0.045 |
| 42.87 | 0.023 | 0.154 | 0.086 | 0.063 | 0.048 | 0.037 | 0.154 | 0.081 | 0.056 | 0.046 | 0.037 |
| 43.88 | 0.020 | 0.136 | 0.076 | 0.055 | 0.042 | 0.033 | 0.136 | 0.071 | 0.049 | 0.040 | 0.033 |
| 44.90 | 0.018 | 0.119 | 0.066 | 0.047 | 0.035 | 0.028 | 0.119 | 0.062 | 0.042 | 0.034 | 0.028 |
| 45.91 | 0.015 | 0.105 | 0.057 | 0.041 | 0.030 | 0.023 | 0.105 | 0.054 | 0.036 | 0.029 | 0.023 |
| 46.92 | 0.013 | 0.093 | 0.050 | 0.036 | 0.027 | 0.020 | 0.093 | 0.047 | 0.032 | 0.025 | 0.020 |
| 47.93 | 0.000 | 0.082 | 0.043 | 0.030 | 0.022 | 0.016 | 0.082 | 0.040 | 0.026 | 0.020 | 0.016 |
| 48.94 | 0.000 | 0.074 | 0.039 | 0.027 | 0.019 | 0.014 | 0.074 | 0.036 | 0.023 | 0.018 | 0.014 |
| 49.95 | 0.000 | 0.066 | 0.034 | 0.023 | 0.016 | 0.012 | 0.066 | 0.032 | 0.020 | 0.015 | 0.012 |

## Table C.2. OneWeb eclipse magnitudes

| HT_KM | Refr | U | B | V | R | I | u | g | r | i | z |
|---|---|---|---|---|---|---|---|---|---|---|---|
| -40.93 | 2.211 | 16.956 | 9.293 | 6.510 | 4.681 | 3.420 | 16.956 | 8.717 | 5.716 | 4.426 | 3.420 |
| -34.20 | 2.117 | 15.907 | 8.604 | 6.014 | 4.335 | 3.191 | 15.907 | 8.065 | 5.283 | 4.103 | 3.191 |
| -16.18 | 1.771 | 10.930 | 6.362 | 4.584 | 3.381 | 2.540 | 10.930 | 6.000 | 4.064 | 3.211 | 2.540 |
| -11.38 | 1.642 | 9.300 | 5.541 | 4.056 | 3.034 | 2.309 | 9.300 | 5.241 | 3.617 | 2.889 | 2.309 |
| -7.11 | 1.515 | 7.842 | 4.786 | 3.556 | 2.698 | 2.082 | 7.842 | 4.538 | 3.189 | 2.575 | 2.082 |
| -3.30 | 1.394 | 6.638 | 4.131 | 3.109 | 2.388 | 1.867 | 6.638 | 3.926 | 2.801 | 2.284 | 1.867 |
| 0.12 | 1.278 | 5.649 | 3.574 | 2.717 | 2.110 | 1.668 | 5.649 | 3.402 | 2.458 | 2.022 | 1.668 |
| 3.28 | 1.163 | 4.801 | 3.086 | 2.367 | 1.856 | 1.483 | 4.801 | 2.942 | 2.150 | 1.782 | 1.483 |
| 6.12 | 1.055 | 4.101 | 2.672 | 2.067 | 1.634 | 1.318 | 4.101 | 2.552 | 1.883 | 1.572 | 1.318 |
| 8.70 | 0.954 | 3.532 | 2.327 | 1.812 | 1.443 | 1.173 | 3.532 | 2.224 | 1.655 | 1.389 | 1.173 |
| 11.05 | 0.858 | 3.067 | 2.036 | 1.594 | 1.277 | 1.045 | 3.067 | 1.948 | 1.459 | 1.231 | 1.045 |
| 13.21 | 0.770 | 2.679 | 1.788 | 1.405 | 1.131 | 0.931 | 2.679 | 1.712 | 1.289 | 1.092 | 0.931 |
| 15.21 | 0.688 | 2.352 | 1.576 | 1.243 | 1.005 | 0.831 | 2.352 | 1.510 | 1.141 | 0.970 | 0.831 |
| 17.06 | 0.612 | 2.078 | 1.395 | 1.103 | 0.894 | 0.743 | 2.078 | 1.337 | 1.014 | 0.864 | 0.743 |
| 18.79 | 0.543 | 1.844 | 1.239 | 0.981 | 0.798 | 0.665 | 1.844 | 1.188 | 0.903 | 0.771 | 0.665 |
| 20.41 | 0.480 | 1.643 | 1.104 | 0.875 | 0.713 | 0.596 | 1.643 | 1.058 | 0.806 | 0.690 | 0.596 |
| 21.94 | 0.424 | 1.470 | 0.986 | 0.782 | 0.638 | 0.535 | 1.470 | 0.945 | 0.720 | 0.618 | 0.535 |
| 23.38 | 0.373 | 1.319 | 0.883 | 0.700 | 0.572 | 0.481 | 1.319 | 0.846 | 0.645 | 0.554 | 0.481 |
| 24.76 | 0.328 | 1.187 | 0.792 | 0.628 | 0.514 | 0.432 | 1.187 | 0.759 | 0.579 | 0.498 | 0.432 |
| 26.08 | 0.287 | 1.069 | 0.711 | 0.563 | 0.461 | 0.389 | 1.069 | 0.681 | 0.520 | 0.447 | 0.389 |
| 27.36 | 0.251 | 0.963 | 0.638 | 0.506 | 0.414 | 0.349 | 0.963 | 0.611 | 0.466 | 0.401 | 0.349 |
| 28.59 | 0.219 | 0.869 | 0.573 | 0.454 | 0.372 | 0.314 | 0.869 | 0.549 | 0.419 | 0.360 | 0.314 |
| 29.79 | 0.191 | 0.784 | 0.515 | 0.407 | 0.334 | 0.282 | 0.784 | 0.493 | 0.376 | 0.323 | 0.282 |
| 30.96 | 0.166 | 0.707 | 0.462 | 0.365 | 0.299 | 0.253 | 0.707 | 0.442 | 0.337 | 0.290 | 0.253 |
| 32.11 | 0.145 | 0.637 | 0.414 | 0.327 | 0.268 | 0.227 | 0.637 | 0.397 | 0.302 | 0.260 | 0.227 |
| 33.24 | 0.125 | 0.573 | 0.371 | 0.293 | 0.240 | 0.203 | 0.573 | 0.355 | 0.270 | 0.233 | 0.203 |
| 34.34 | 0.109 | 0.516 | 0.332 | 0.262 | 0.214 | 0.182 | 0.516 | 0.318 | 0.241 | 0.208 | 0.182 |
| 35.44 | 0.094 | 0.464 | 0.297 | 0.234 | 0.191 | 0.162 | 0.464 | 0.284 | 0.215 | 0.185 | 0.162 |
| 36.52 | 0.081 | 0.416 | 0.265 | 0.208 | 0.171 | 0.145 | 0.416 | 0.254 | 0.192 | 0.165 | 0.145 |
| 37.59 | 0.070 | 0.373 | 0.237 | 0.185 | 0.152 | 0.129 | 0.373 | 0.226 | 0.171 | 0.147 | 0.129 |
| 38.64 | 0.061 | 0.334 | 0.211 | 0.165 | 0.135 | 0.114 | 0.334 | 0.201 | 0.152 | 0.131 | 0.114 |
| 39.69 | 0.053 | 0.299 | 0.188 | 0.147 | 0.120 | 0.102 | 0.299 | 0.179 | 0.135 | 0.116 | 0.102 |
| 40.73 | 0.045 | 0.268 | 0.167 | 0.130 | 0.106 | 0.090 | 0.268 | 0.159 | 0.120 | 0.103 | 0.090 |
| 41.77 | 0.039 | 0.239 | 0.148 | 0.116 | 0.094 | 0.080 | 0.239 | 0.142 | 0.106 | 0.091 | 0.080 |
| 42.80 | 0.034 | 0.214 | 0.132 | 0.103 | 0.084 | 0.071 | 0.214 | 0.126 | 0.094 | 0.081 | 0.071 |
| 43.82 | 0.029 | 0.191 | 0.118 | 0.091 | 0.074 | 0.063 | 0.191 | 0.112 | 0.084 | 0.072 | 0.063 |
| 44.85 | 0.025 | 0.171 | 0.105 | 0.081 | 0.066 | 0.056 | 0.171 | 0.100 | 0.074 | 0.064 | 0.056 |
| 45.86 | 0.022 | 0.153 | 0.093 | 0.072 | 0.058 | 0.049 | 0.153 | 0.089 | 0.066 | 0.056 | 0.049 |
| 46.88 | 0.019 | 0.138 | 0.083 | 0.064 | 0.052 | 0.044 | 0.138 | 0.079 | 0.059 | 0.050 | 0.044 |
| 47.90 | 0.017 | 0.124 | 0.075 | 0.058 | 0.047 | 0.039 | 0.124 | 0.071 | 0.053 | 0.045 | 0.039 |
| 48.91 | 0.014 | 0.112 | 0.067 | 0.052 | 0.042 | 0.035 | 0.112 | 0.064 | 0.047 | 0.040 | 0.035 |
| 49.92 | 0.012 | 0.101 | 0.060 | 0.046 | 0.037 | 0.031 | 0.101 | 0.057 | 0.042 | 0.036 | 0.031 |

## Appendix D. Tracks of observations used for comparison with the model

Magnitudes from the following eight MMT-9 tracks were used for the comparison with the model in Section 7: 18801226, 19733459, 18362701, 19612472, 19344200, 18203200, 18903629 and 19956408. Two tracks, 19339509 and 19344243, were omitted because of large brightness surges during the eclipse. The light curves for the omitted passes are shown in Figure D-1.

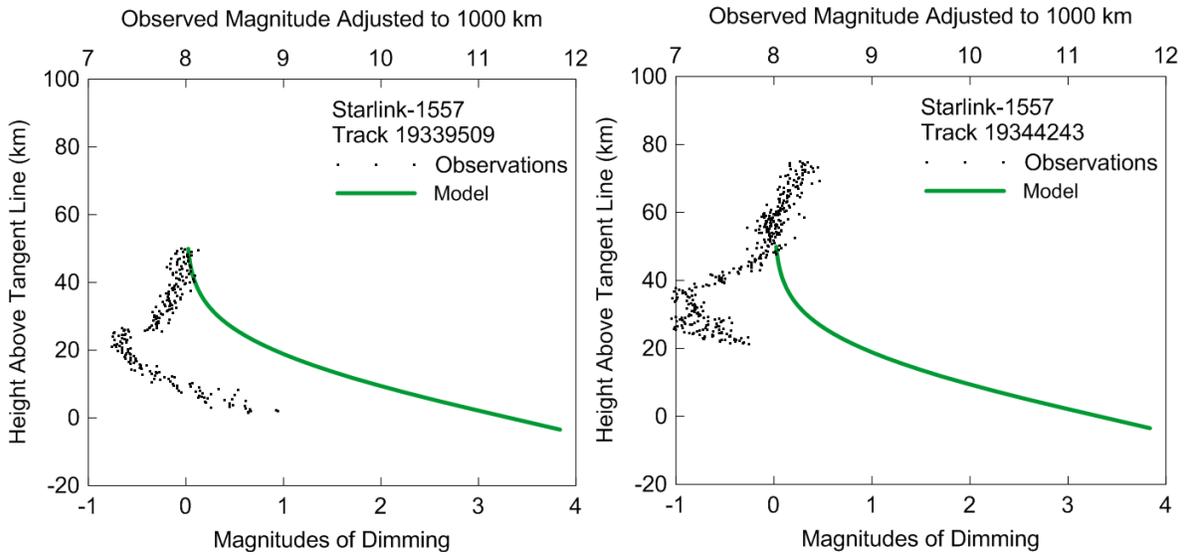

Figure D-1. These tracks of observations were omitted from the comparison with the eclipse model described in Section 7 because large brightness surges occurred during the eclipse. The light curves after the surges show that the slope of observed brightness agrees with that of the model.